\documentclass[useAMS,usenatbib, usegraphicx]{mn2e}
\title[Spin-resolved spectroscopy of DQ~Her]{Spin-resolved spectroscopy of the intermediate polar DQ~Her}

\author[S. Bloemen et al.]{S. Bloemen$^{1}$\thanks{E-mail:
steven.bloemen@ster.kuleuven.be}, T. R. Marsh$^{2}$, D. Steeghs$^{2,3}$ and R. H. \O stensen$^{1}$\\
$^{1}$Instituut voor Sterrenkunde, Katholieke Universiteit Leuven, Celestijnenlaan 200D, B-3001 Leuven, Belgium\\
$^{2}$Department of Physics, University of Warwick, Coventry CV4 7AL, UK\\
$^{3}$Harvard-Smithsonian Center for Astrophysics, 60 Garden Street, Cambridge, MA 02138, USA}
\begin{document}

\date{Accepted 2010 May 17. Received 2010 May 17; in original form 2010 February 26}

\pagerange{\pageref{firstpage}--\pageref{lastpage}} \pubyear{2010}

\maketitle

\label{firstpage}

\begin{abstract}
We present high-speed spectroscopic observations of the intermediate polar DQ~Herculis. Doppler tomography of two He~I lines reveals a spiral density structure in the accretion disc around the white dwarf primary. The spirals look very similar to the spirals seen in dwarf novae during outburst. DQ~Her is the first well established intermediate polar in which spirals are seen, that are in addition likely persistent because of the system's high mass transfer rate. Spiral structures give an alternative explanation for sidebands of the WD spin frequency that are found in IP light curves. The Doppler tomogram of He~II $\lambda4686$ indicates that a large part of the emission is not disc-like.

Spin trails of spectra reveal a pulsation in the He~II $\lambda4686$ emission that is believed to result from reprocessing of X-rays from the white dwarf's magnetic poles in the accretion flow close to the WD. We confirm the previous finding that the pulsation is only visible in the red-shifted part of the line when the beam points to the back side of the disc. The absence of reprocessed light from the front side of the disc can be explained by obscuration by the front rim of the disc, but the absence of extra emission from the blue-shifted back side of the disc is puzzling. Reprocessing in accretion curtains can be an answer to the problem and can also explain the highly non-Keplerian velocity components that are found in the He~II $\lambda4686$ line. Our spin trails can form a strong test for future accretion curtain models, with the possibility of distinguishing between a spin period of 71\,s or 142\,s. Spin trails of data taken at selected orbital phases show little evidence for a significant contribution of the bright spot to the pulsations and allow us to exclude a recent suggestion that 71\,s is the beat period and 70.8\,s the spin period.
\end{abstract}



\section[]{Introduction}\label{sec_intro}
DQ~Her is the prototype of the intermediate polars, which are therefore also called `DQ~Herculis stars'. Intermediate polars are cataclysmic variables with moderately magnetic white dwarfs. The magnetic field in such systems is not strong enough to synchronise the binary orbit and the spin period of the white dwarf (contrary to the systems with higher magnetic fields, called polars) but it disrupts the inner part of the accretion disc. The disc material finally accretes on the magnetic poles of the white dwarf via the magnetic field lines. The gravitational energy that gets released is radiated in X-rays, which results in X-ray beams from the magnetic poles of the white dwarf. Reprocessing of X-rays from the beams in the accretion disc or on the surface of the secondary star leads to periodicities in photometry and spectra, which are a very important characteristic of IPs. An excellent review on intermediate polars can be found in \citet{Patterson1994}.

\citet{Walker1954, Walker1956} first reported a periodic 71\,s variation in DQ~Her's light curve. Spectroscopic studies revealed that a disc is present in the system \citep{GreensteinKraft1959}. \cite{WarnerPeters1972} found a phase shift in the 71\,s pulsations during eclipse, which was later confirmed \citep[e.g.,][]{PattersonRobinson1978,ZhangRobinson1995}. Several studies \citep[see e.g.,][]{ChananNelson1978, Chester1979, Petterson1980} showed that this phase shift is naturally explained by pulsations in the light curve that arise from illumination of the accretion disc by energetic beams from the white dwarf with the front side of the disc hidden behind the disc rim.

For decades a debate has gone on as to whether the true spin period of the white dwarf is the observed 71\,s or 142\,s. In the latter case, the 71\,s periodicity that is observed could be the result of two beams illuminating the disc with a 0.5 phase difference. In power spectra of photometric data, usually no power is found at 142\,s \citep[e.g.,][]{KiplingerNather1975,WoodRobertson2005}. The radiation beams from the two magnetic poles of the white dwarf should thus be identical to a very high approximation, which is not impossible but at least physically difficult to achieve. Weak photometric evidence of a 142\,s periodicity was, however, presented in \citet{Nelson1975} and \citet{SchoembsRebhan1989}. Furthermore, the models for the phase shift that were presented in the 1970s and 1980s did not rule out a 142\,s spin period. \citet{ZhangRobinson1995} claimed to have found a better agreement with their observations for a 142\,s model than for a 71\,s model. In polarimetric data, \citet{KempSwedlund1974} and \citet{SwedlundKemp1974} found a stronger 142\,s than 71\,s periodicity. In the polarimetric study by \citet{ButtersKatajainen2009}, only the 142\,s periodicity was searched for (and detected but with low significance) because the time resolution and signal-to-noise ratio of the dataset were not high enough to allow a study of the 71\,s periodicity (Butters 2009, private communication). Spectroscopy has so far supported both the 71\,s and 142\,s spin periods \citep{ChananNelson1978,MartellHorne1995}. All things considered, it is not clear yet which of the spin periods is to be preferred. 

DQ~Her is eclipsing, with an inclination of $\approx 89^\circ$ \citep{Petterson1980}. It has an orbital period of 4h39m and consists of a white dwarf with a mass near 0.6 $\rm{M}_\odot$ and a red dwarf companion with a mass near 0.4 $\rm{M}_\odot$ \citep{HorneWelsh1993}. 

This paper starts with a discussion on the observations and the data reduction process in Section~\ref{sec_obs}. In Section~\ref{sec_DM}, the variations in the spectral lines on the timescale of the binary orbit are studied using Doppler tomography to get more insight in the binary's structure. Next, in Section~\ref{sec_spin}, the dependance on the spin phase is analysed in an attempt to get clues about the true spin period and the X-ray reprocessing regions. Finally, the conclusions are given in Section~\ref{sec_concl}.


\section[]{Observations and data reduction} \label{sec_obs}
Spin resolved spectroscopy of DQ~Her was performed with the 4.2-m William Herschel Telescope on La Palma (Spain), on the nights of 8-10 July 1998. Approximately 3 orbital periods were observed using the double-armed ISIS spectrograph. The CCDs were operated in low smear drift mode \citep{RuttenGribbin1997} to limit the dead-time between two exposures to about 0.6\,s. The blue arm spectra cover the spectral range $4200-5000$\,\AA\ and have a spectral resolution of 1.9\,\AA. The spectral range covered by the red arm spectra is $6320-6710$\,\AA\ and their spectral resolution is 0.8\,\AA. An overview of the observations is given in Table \ref{tab_obs}. The observing conditions were generally good with a few cirrus clouds at times and some dust in the air during the first night.

\begin{table*}
 \centering
 \begin{minipage}{139mm}
  \caption{Overview of our high-speed spectroscopic observations of DQ~Her, performed with the William Herschel Telescope.}
  \begin{tabular}{@{}llll lrr@{}}
  \hline
      Date & UT & Seeing & Instrument & Grating & Exp. time & Number\\
      \hline
      08-09/07/1998 & 22:32 - 04:19 &  $\approx 1.2$ arcsec & ISIS blue arm & R1200B & 5.1s & 2419 \\
      	\vspace{0.1cm}
                              &  & & ISIS red arm & R600R & 15.1s & 908 \\
      09-10/07/1998 & 22:48 - 05:25 & $1$ to $2$ arcsec & ISIS blue arm & R1200B & 5.1s & 3181 \\
	\vspace{0.1cm}
                              & & & ISIS red arm & R600R & 15.1s & 1170 \\
      10-11/07/1998 & 00:06 - 04:35 & $\approx 1$ arcsec but worse & ISIS blue arm & R1200B & 5.1s & 1792 \\
                              & &  during last hour & ISIS red arm & R600R & 15.1s & 780 \\
\hline
\label{tab_obs}
\end{tabular}
\end{minipage}
\end{table*}

One-dimensional spectra were extracted from the two-dimensional frames using \texttt{STARLINK}\footnote{\texttt{STARLINK} is open source software and can be obtained from http://starlink.jach.hawaii.edu/starlink .} routines and the \texttt{PAMELA} and \texttt{MOLLY}\footnote{\texttt{PAMELA}, \texttt{MOLLY} and \texttt{DOPPLER} are available for download at http://deneb.astro.warwick.ac.uk/phsaap/software/ .} software packages written by TRM. Each frame was flatfielded using sky flats to remove the spatial structure and lamp flats to correct for the CCD's pixel-to-pixel sensitivity differences. The sky contribution was removed by subtracting a polynomial fit to the sky regions of each row of the CCD frame in spatial direction. Finally, the optimal extraction method described in \citet{Horne1986} was used to extract one-dimensional spectra. The arc spectra were wavelength calibrated by a polynomial fit to the lines. The DQ~Her spectra were wavelength calibrated by linear interpolation of the wavelength scales of an arc spectrum taken before and after the science spectrum. 

There was no comparison star available due to the limited CCD area that can be used in low smear drift mode. The spectra could therefore not be flux-calibrated. Because of seeing variations and variable slit losses, our ability to study changes in the continuum level of the spectra was very limited. Therefore, the continuum level was subtracted such that variations in the emission lines could be studied.


\section[]{Doppler maps} \label{sec_DM}

\begin{figure}
\includegraphics[width=84mm]{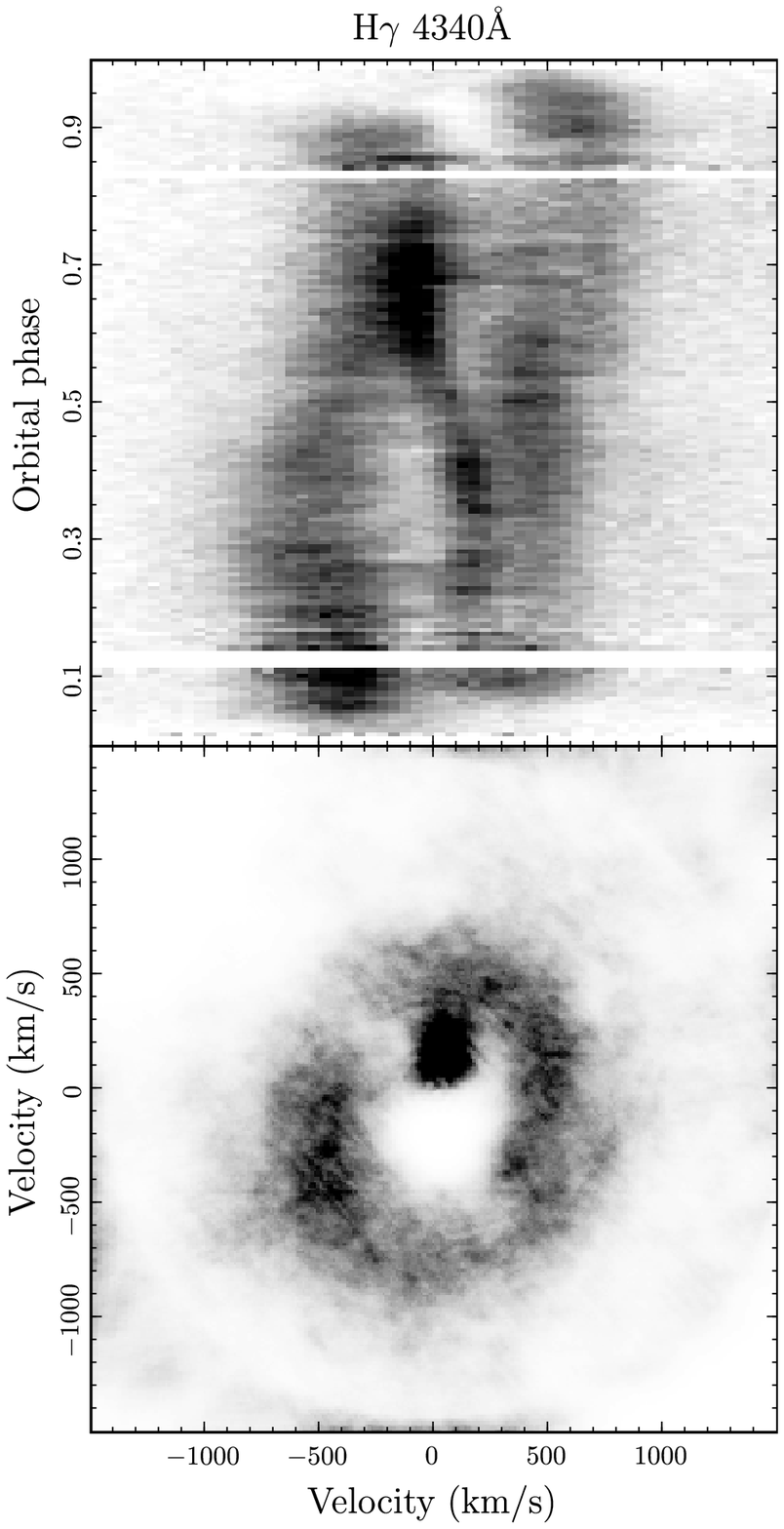}
 \caption{The observed H$\gamma$ flux of DQ~Her in function of orbital phase (top panel) and the corresponding Doppler map (bottom panel). The continuum level is set to white. The accretion disc and the secondary star show up on the Doppler map, as expected for a cataclysmic variable.}
  \label{DM_4340}
\end{figure}

\begin{figure}
\includegraphics[width=84mm]{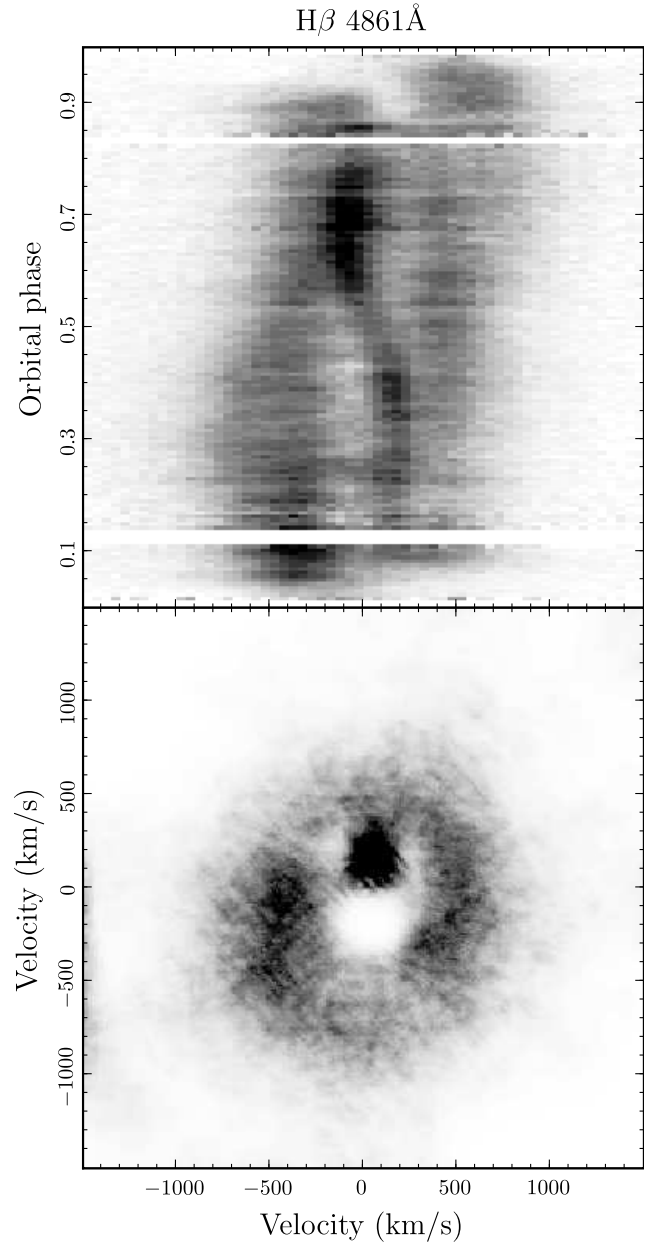}
 \caption{The observed H$\beta$ flux of DQ~Her in function of orbital phase (top panel) and the corresponding Doppler map (bottom panel). The Doppler map shows the same components as the H$\gamma$ map in Fig.~\ref{DM_4340}. }
  \label{DM_4861}
\end{figure}

\begin{figure}
\includegraphics[width=84mm]{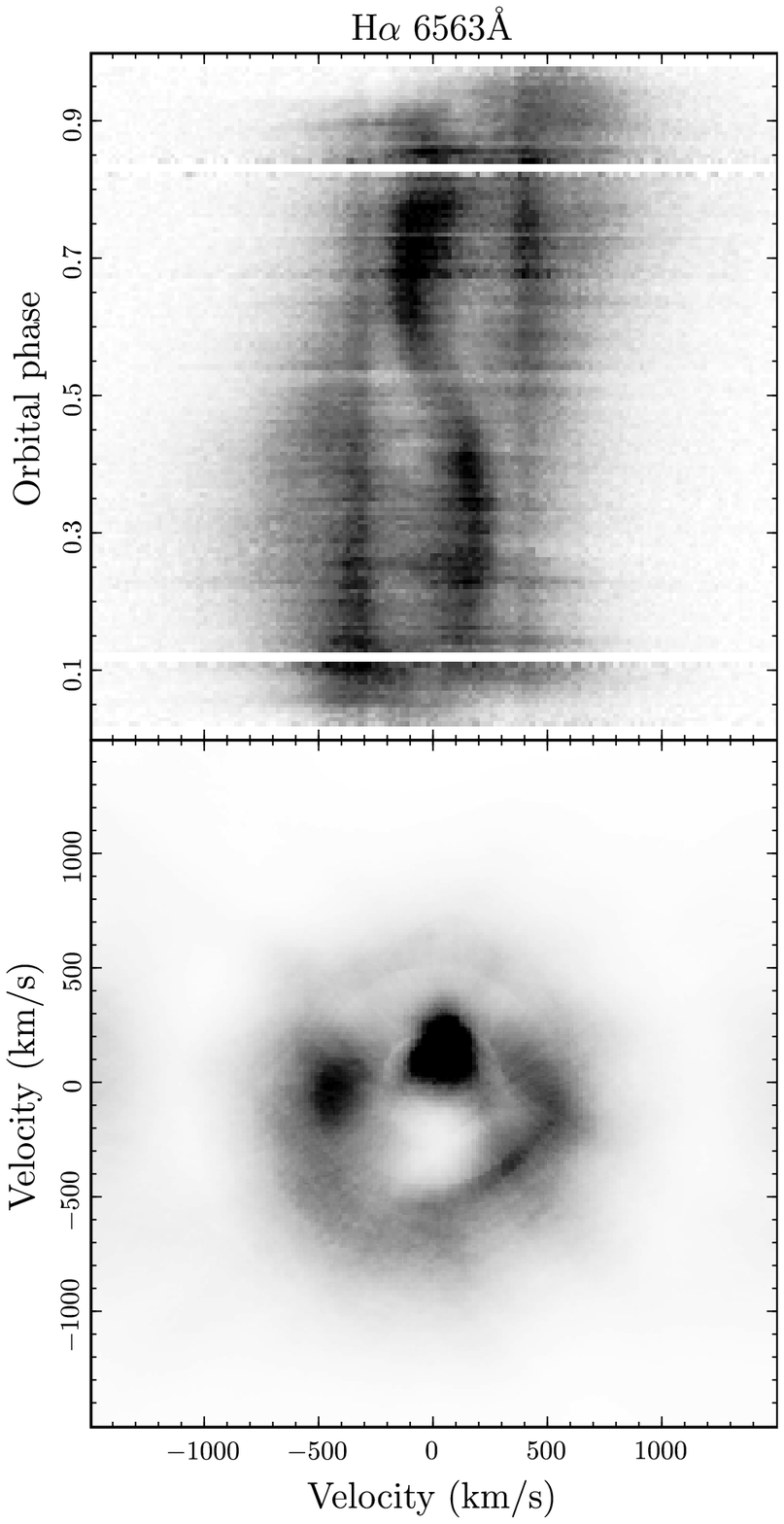}
 \caption{The observed H$\alpha$ flux of DQ~Her in function of orbital phase (top panel) and the corresponding Doppler map (bottom panel). Emission from the accretion disc and the secondary is present as in the other Balmer lines, but additional stationary components are observed at $-388\pm2$ and $+337\pm2$ km s$^{-1}$ which are probably emitted by DQ~Her's nova shell. A region of 160 km~s$^{-1}$ around these components was masked when creating the Doppler image.}
  \label{DM_6563}
\end{figure}

We first study the dependence of the spectral lines on the orbital phase of the binary. The orbital phases of the spectra were calculated using the quadratic ephemeris of \citet{WoodRobertson2005}. The continuum subtracted spectra were phasebinned and rebinned to a constant velocity scale of 50 km s$^{-1}$ per pixel for the blue arm spectra and 18 km s$^{-1}$ for the red arm spectra. The top panels of Figs.~\ref{DM_4340} to \ref{DM_4686} show the resulting orbital trails for selected emission lines. Blank horizontal lines indicate uncovered orbital phases. To shed light on the binary components that contribute to the phase-dependent line profiles, Doppler tomograms have been produced, which are shown on the bottom panels of the same figures. 

Line profiles that are observed at different orbital phases are effectively projections of a rotating velocity image of the binary. Doppler tomography allows the construction of the velocity image from observed spectra. The images shown here have been created using the maximum entropy method (MEM) as implemented in TRM's \texttt{DOPPLER} package\footnotemark[\value{footnote}], which is an iterative procedure to build up an image that reproduces the observed profiles best while striving towards a map of least structure. A detailed description of Doppler tomography using MEM is presented in \citet{MarshHorne1988}.

Spectra taken between orbital phases -0.1 and 0.1 have not been used to avoid fitting the effect of the eclipse. A systemic velocity of $\gamma=-60$ km s$^{-1}$ was adopted (\citealt{HutchingsCowley1979}).


\subsection{Balmer lines}\label{sec_DM_balmer}

The tomograms of H$\gamma$ and H$\beta$ (Figs.~\ref{DM_4340} and \ref{DM_4861}) look like the velocity images of a  typical cataclysmic variable. The broad line wings that are emitted by the accretion disc around the white dwarf map onto the diffuse ring on the Doppler maps. The ring is shown inside out because the disc components with the highest velocities are the ones closest to the white dwarf in positional space. The S-wave on the trail is emission from the secondary star and maps onto a bright dot around $K_2$, which \citet{HorneWelsh1993} determined to be 227~km s$^{-1}$. On the trails of the spectra, the rotational disturbance of the line through eclipse is clearly visible \citep{YoungSchneider1980}. The blue-shifted part of the accretion disc emission gets eclipsed first and reappears earlier than the red-shifted part.

The H$\alpha$ line (Fig.~\ref{DM_6563}) is found to have stationary components at $-388\pm2$ and $+337\pm2$ km s$^{-1}$. These components were already reported by \citet{BianchiniMastrantonio2004}. Contrary to their findings, however, we do see emission from the nova shell in the `sky regions' of our 2D spectra and therefore believe that the stationary emission components are emitted by this shell. The shell is the result of the nova which took place in 1934. \citet{VaytetOBrien2007} estimated the expansion velocity of the nova at 370 km s$^{-1}$, which looks to be consistent with our findings.
Doppler tomography assumes that all light is emitted by material that corotates with the binary. When creating the Doppler map for H$\alpha$, a region of 160 km s$^{-1}$ around the stationary components was masked to give the best possible representation of the structure of the binary system, but this masking also introduces uncertainties in a ring with a width of $\approx 160$ km~s$^{-1}$ and a radius of $\approx 350$ km s$^{-1}$ from the center of the map.


\subsection{Spiral structure on the He~I $\lambda4472$ and $\lambda6678$ maps}\label{sec_DM_HeI}

\begin{figure}
\includegraphics[width=84mm]{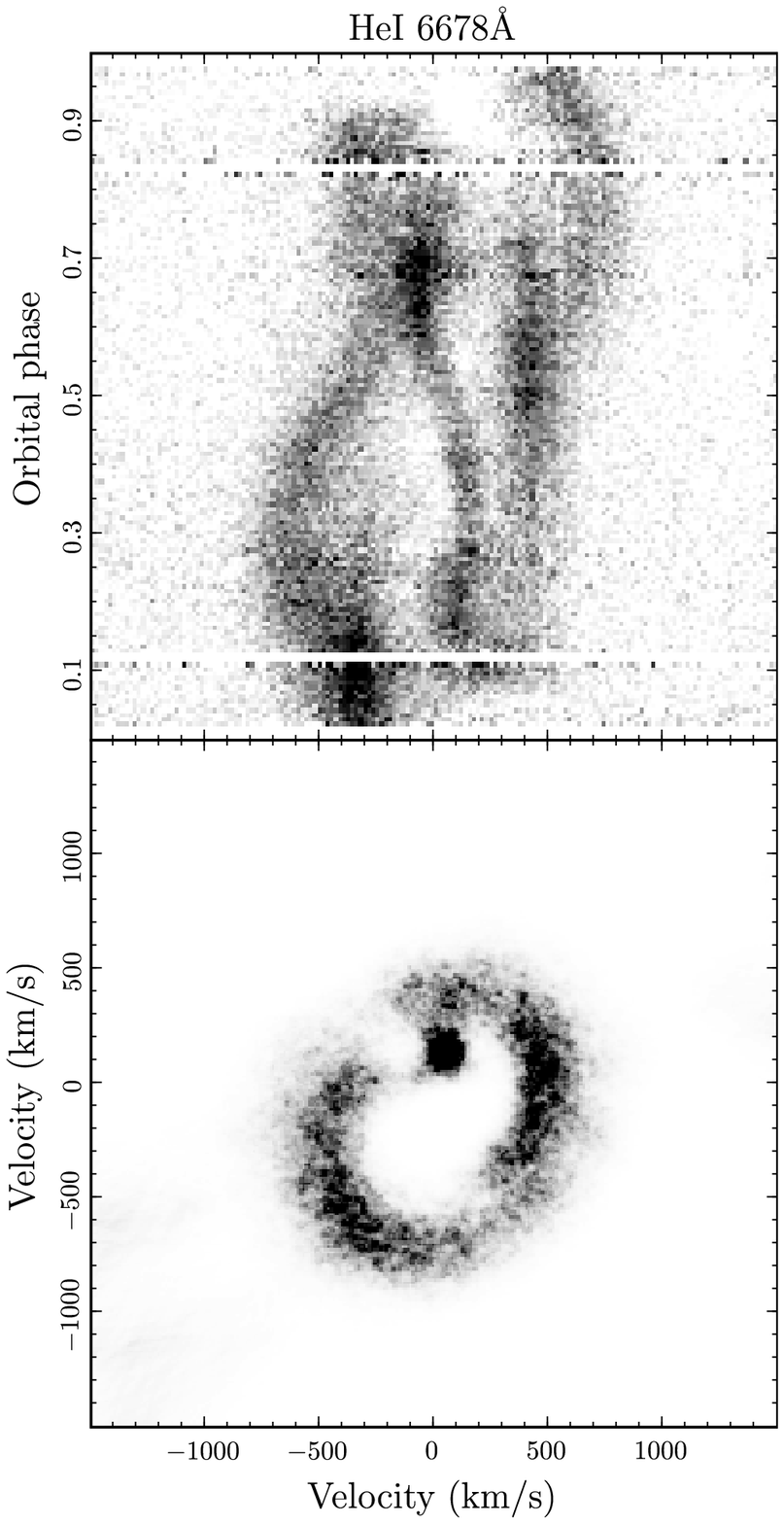}
 \caption{The observed He~I $\lambda6678$ flux of DQ~Her in function of orbital phase (top panel) and the corresponding Doppler map (bottom panel). Spiral arms are detected in the accretion disc, which are caused by tidal effects from the secondary on the outer part of the disc.}
  \label{DM_6678}
\end{figure}

\begin{figure}
\includegraphics[width=84mm]{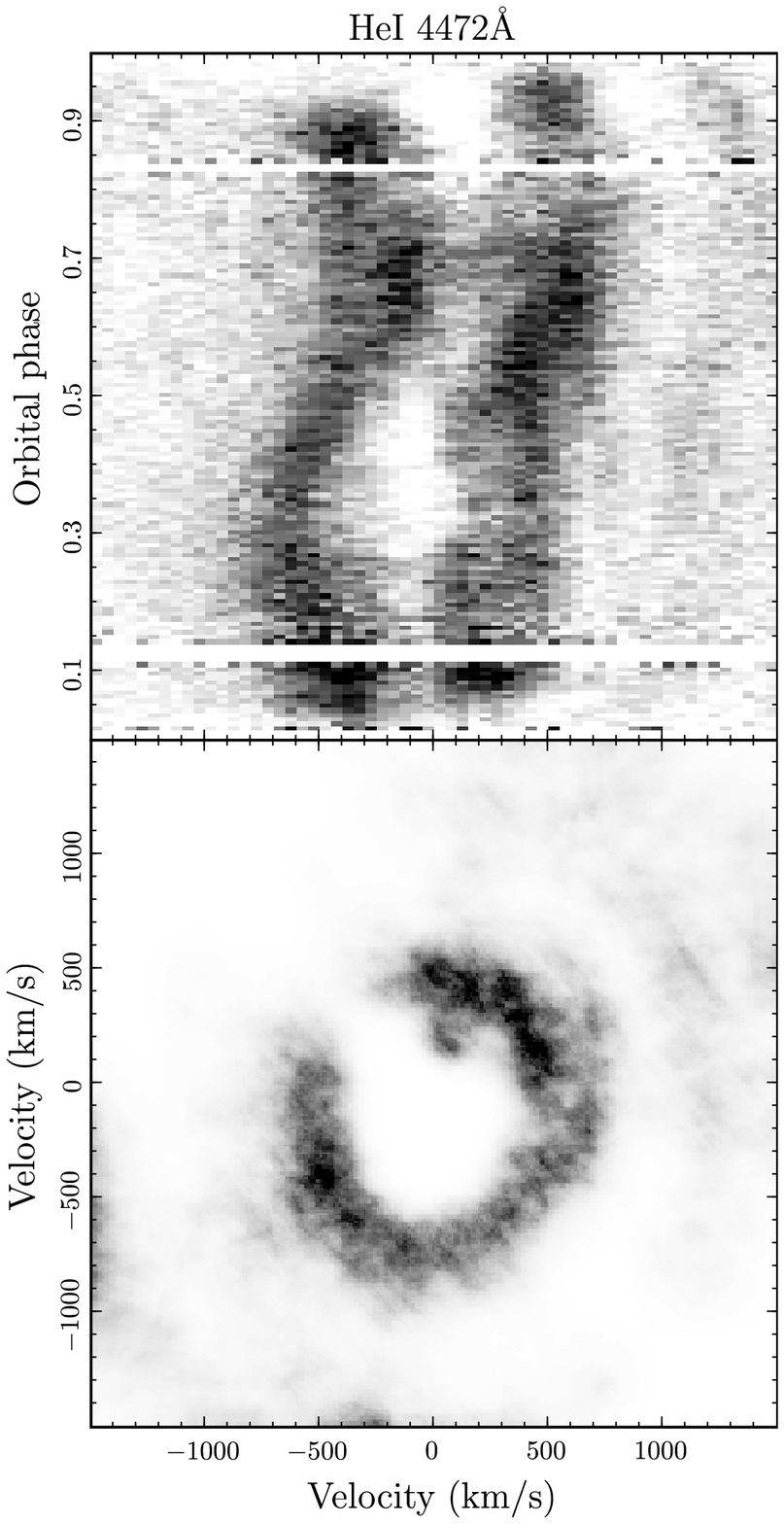}
 \caption{The observed He~I $\lambda4472$ flux of DQ~Her in function of orbital phase (top panel) and the corresponding Doppler map (bottom panel). As in the He~I $\lambda6678$ map (Fig.~\ref{DM_6678}), spiral arms are seen in the accretion disc.}
  \label{DM_4472}
\end{figure}

The He~I maps (Figs.~\ref{DM_6678} and \ref{DM_4472}) reveal spiral structure in the disc, which is also present on the H$\gamma$ tomogram. The spiral pattern is comparable to that found in a number of dwarf novae during outburst and in novalike CVs. 
Spirals were first discovered in the dwarf nova IP Peg in outburst by \citet{SteeghsHarlaftis1997}. In that paper it was shown that spirals in velocity space map onto spirals in positional coordinates as well. A review of observations of spirals in CV discs can be found in \citet{Steeghs2001}. Spirals in accretion discs of CVs are believed to arise from tidal forces from the secondary on the outer regions of the disc. Because the material in the disc gradually spirals inwards to orbits with a higher Keplerian velocity, a double spiral arm is created. 

Systems in which spirals are seen, have a high mass transfer rate and thus a large accretion disc, filling most of the Roche lobe of the white dwarf. This is necessary for the secondary to have a large enough tidal effect \citep{SavonijePapaloizou1994}. The disc of DQ~Her extends up to at least 87\% of the white dwarf's Roche lobe (\citealt{Harrop-AllinWarner1996}, determined from the eclipse duration). In dwarf novae, the spirals are only observed during outburst when the mass transfer rate is temporarily increased, i.e.\ for only a few days in an outburst cycle of the order of months. This makes them particularly difficult to observe because the outburst moments are difficult to predict. 

This is the first detection of spirals in the accretion disc of a well-established intermediate polar. The fact that the WD is magnetic does not seem to preclude spirals. Given that their formation is driven by tides in the outer disc and does not directly depend on the magnetically-controlled inner disc, this does not come as a surprise. In WZ Sge, which is possibly an IP, spirals have been detected as well \citep{BabaSadakane2002} but only during outburst. 

Several systems have been proposed to explain the apparent presence of spiral structure in accretion discs \citep[see e.g.,][]{SteeghsStehle1999, Smak2001, Ogilvie2002}. Our data do not allow to descriminate between the different explanations, but DQ~Her is a good candidate system for further research on the spirals in accretion discs. Since the mass transfer in DQ~Her is believed to be in an equilibrium state, the spiral structure is very likely to be permanently visible, which is a huge advantage over the known dwarf novae which only show spirals during outbursts.


\subsection{He~II $\lambda4686$} \label{sec_DM_HeII}

\begin{figure}
\includegraphics[width=84mm]{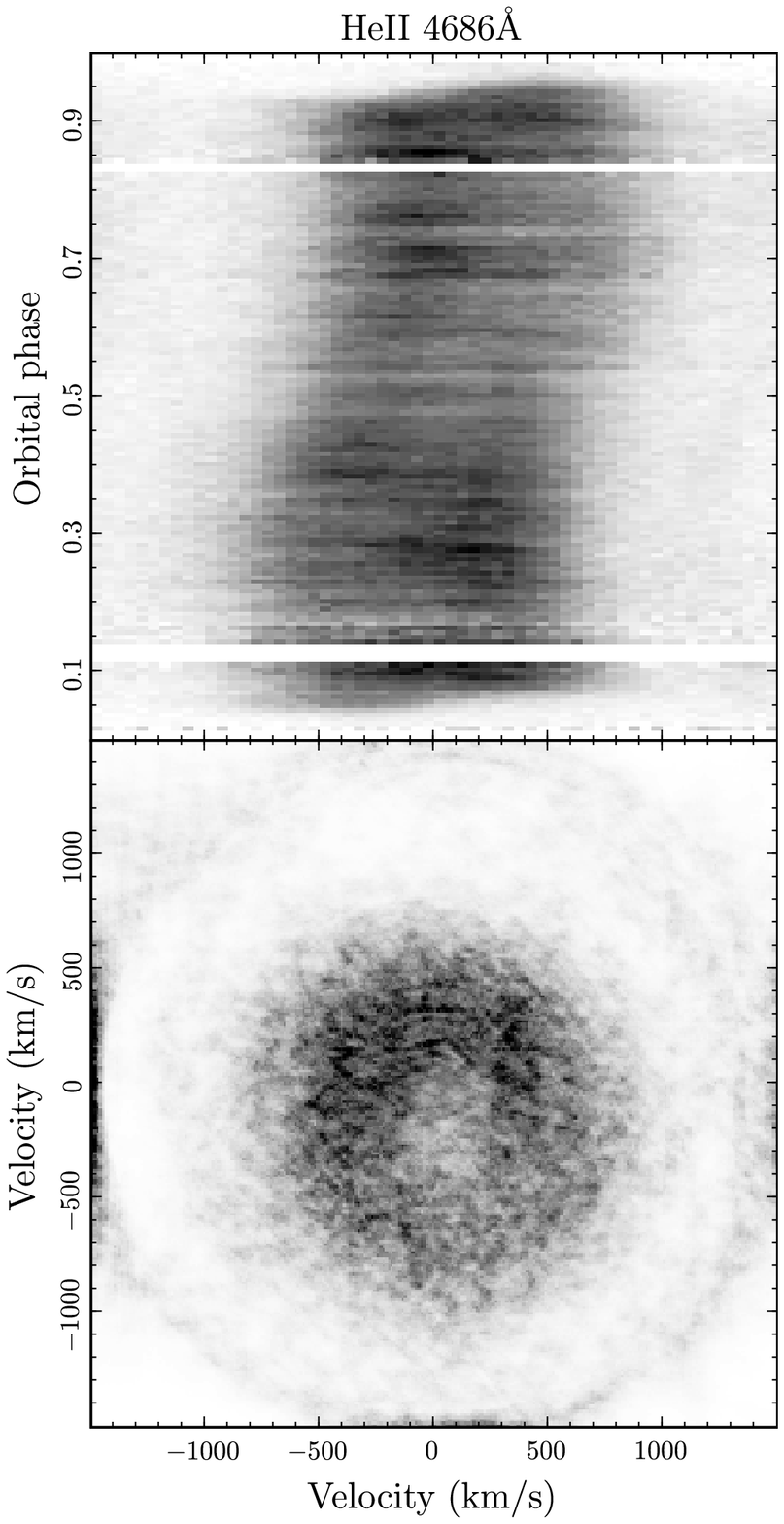}
 \caption{The observed He~II $\lambda4686$ flux of DQ~Her in function of orbital phase (top panel) and the corresponding Doppler tomography image (bottom panel). The very diffuse disc probably indicates that a substantial part of the line emitting material has a strong non-Keplerian velocity component.}
  \label{DM_4686}
\end{figure}

The map of He~II $\lambda4686$ (Fig.~\ref{DM_4686}) shows a filled ring. As \citet{MartellHorne1995} already pointed out, this reflects the absence of a double peaked line profile. The He~II $\lambda4686$ line is formed after electron capture by He~III. The ionisation energy of He~I is 24.6\,eV and further ionisation of He~II requires 54.4\,eV. High temperatures are needed to make this possible, which are only found in the inner regions of the disc. In this region, the material has high Keplerian velocities. The presence of a lot of emission at lower velocities shows that the line has a much higher non-Keplerian component than the other lines. We therefore consider it very likely that a substantial part of the line is produced in a region that is influenced by the magnetic field, because material that gets trapped by magnetic field lines gets a larger than usual velocity component towards the white dwarf as well as a non-negligible component in the direction perpendicular to the disc. We can also confirm the observation by \citet{MartellHorne1995} of high emission regions on the upper half of the map. The higher emission in the disc part $V_X<0, V_Y>0$ can be associated with the accretion stream, but the high emission at $V_X>0$ is not easily explained.\\

In an attempt to find the variations in the line profiles due to reprocessed X-rays of the white dwarf in the disc or on the secondary star, Doppler maps were also created for different phases in the spin period and the beat period, but no significant variations were found.


\subsection{Discussion}
Angular momentum transport in accretion discs is still actively studied. The famous ``$\alpha$-description'' by \citet{ShakuraSunyaev1973} explains the momentum transport by a viscosity effect. The effective viscosity is believed to be driven by MHD instabilities that make shear flows turbulent (for an extensive review see \citealt{Balbus2003}). Spiral arms are of notable astrophysical interest because spiral shocks are believed to be an alternative (or complementary) means to transfer angular momentum. A review of the theoretical efforts on spiral shocks is presented in \citet{Boffin2001}. The establishment of spiral waves in an accretion disc of an IP also re-enforces the findings by \citet{MurrayArmitage1999} that tidally induced spirals can propagate sufficiently far into the disc of an IP such that they can modulate the accretion rate onto the white dwarf. This could explain the sidebands of the spin frequency that are often found in X-ray and optical light curves of IPs. These sidebands were thought to be a sign of mass accretion via a direct mass stream from the first Lagrange point to the white dwarf rather than via an accretion disc and have been used to discriminate between systems that have only disc accretion, only stream accretion or a mixture of both (see e.g.,\ \citealt{NortonBeardmore1996} for a model of the X-ray power spectrum and \citealt{Hellier2007} for a review on accretion in IPs). If spirals can cause the sidebands as well, the conclusions drawn on this basis may require reassessment.


\section[]{Spin phase dependance of the spectral lines} \label{sec_spin}

Our high-speed spectroscopic dataset also allowed us to study the variations in the spectral line shapes on the spin period of the binary's white dwarf. X-rays that are emitted from the white dwarfs' magnetic poles in intermediate polars can make the spin period  of the white dwarf visible in light curves and spectra of an IP. As the white dwarf spins, the X-ray beams sweep around, creating a light house effect. If such a beam (or both beams) illuminate the accretion disc, reprocessing of the X-rays can lead to enhanced continuum and line emission. As mentioned in Section \ref{sec_intro}, in the case of DQ~Her, depending on the study a 71\,s or 142\,s periodicity is found. This periodicity is too stable to be an asteroseismological effect and is therefore very likely to be associated with the spin frequency of the white dwarf. A less favoured explanation is that the 71\,s periodicity is in fact the beat period ($\omega_{beat}=\omega_{spin}-\omega_{orb}$), which would be the case if the vast majority of reprocessing is done in the bright spot where the accretion stream hits the disc (e.g.,\ \citealt{SaitoBaptista2009}) or on the surface of the secondary star, since both orbit at the orbital frequency.


\subsection{Spin periodicity in the red-shifted wing of He~II $\lambda4686$}

\begin{figure}
\includegraphics[width=84mm]{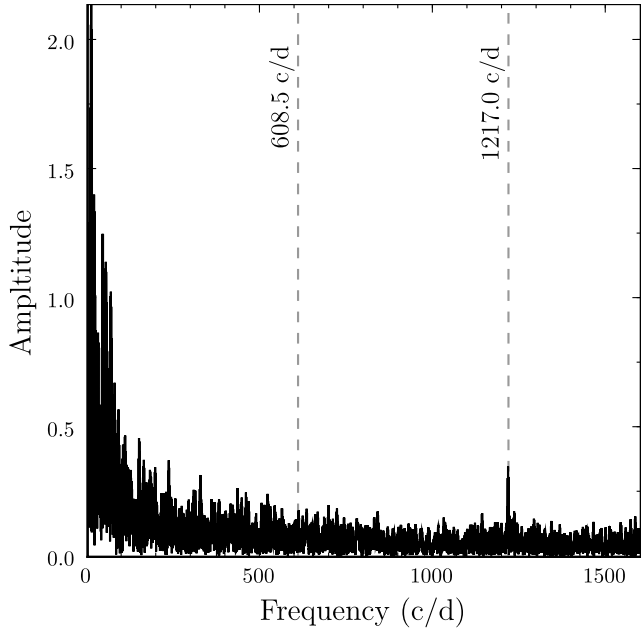}
 \caption{Periodogram of the flux in the He~II $\lambda4686$ line. The 71\,s periodicity (1217.0 c\,d$^{-1}$) is retrieved, but no 142\,s periodicity (608.5 c\,d$^{-1}$) is found.}
  \label{SCARGLE_4686}
\end{figure}

Since our spectra were taken without a comparison star in the slit, we could not correct for variations in the observed intensity that are due to atmospheric conditions and slit losses. We thus worked with continuum-subtracted spectra and limited our study of the oscillations to the flux in the emission lines. These can fluctuate more than the continuum level if the high energetic UV and X-ray photons of the WD beams trigger extra line transitions. A Lomb-Scargle periodogram \citep{Lomb1976,Scargle1982} for the He~II $\lambda4686$ line flux is shown in Fig.~\ref{SCARGLE_4686}. A peak is found at 1217 cycles per day (c\,d$^{-1}$), which is equivalent to a period of 71\,s. No pulsation amplitude is found at the 142\,s period (608.5 c\,d$^{-1}$). This result is in line with the outcome of most photometric studies which find the white dwarf spin period of 71\,s, as described in Section~\ref{sec_intro}. Periodograms of the fluxes in the Balmer and He~I lines that were used in Sections \ref{sec_DM_balmer} and \ref{sec_DM_HeI} show no frequencies above noise level (not shown). The line transition in which these photons are emitted is thus not strongly influenced by the beam photons or the emitting region is not easily reached by the beam. 

To visualise the effect of the X-ray reprocessing on the spectral line flux, we followed the same approach as \citet{MartellHorne1995}. We folded the data on the 71\,s and 142\,s periods and subtracted a mean spectrum. This way, only the deviations from the mean value are left. Only spectra taken at orbital phases from 0.1 up to 0.9 are used, i.e.\ spectra taken during eclipse are omitted. 

\begin{figure}
\includegraphics[width=84mm]{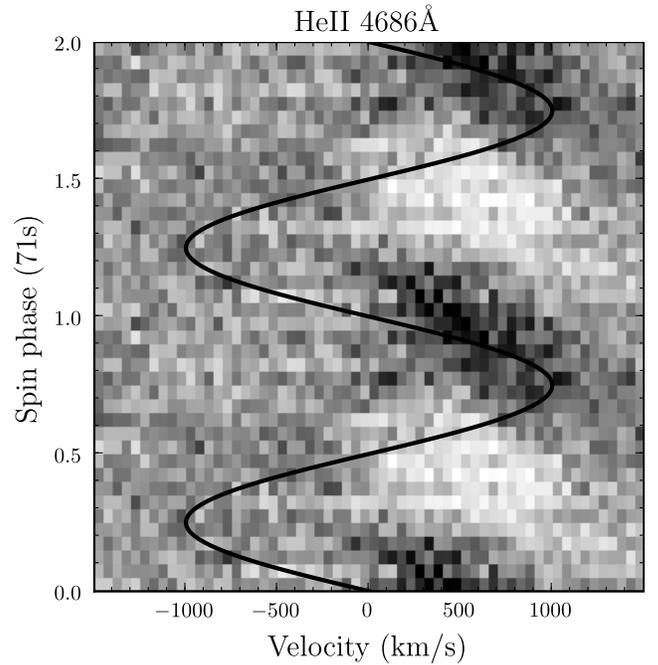}
 \caption{Trail of continuum and mean subtracted He~II $\lambda4686$ line profiles, folded on a 71\,s period. The spin phase is shown twice to ease comparison with the next figure. In the blue-shifted part of the line, no modulation is seen on the 71\,s period. In the red-shifted part, however, enhanced emission is seen once per phase (darker region). Note that the white region does not denote a lower emission than the grey region in the blue-shifted part, but a lower emission than the average on those particular wavelengths. The black sinusoidal curve shows where enhanced emission from reprocessing of the light of one white dwarf X-ray beam in a small part of the disc would be seen.}
  \label{SPIN_4686_71s_rep2}
\end{figure}

\begin{figure}
\includegraphics[width=84mm]{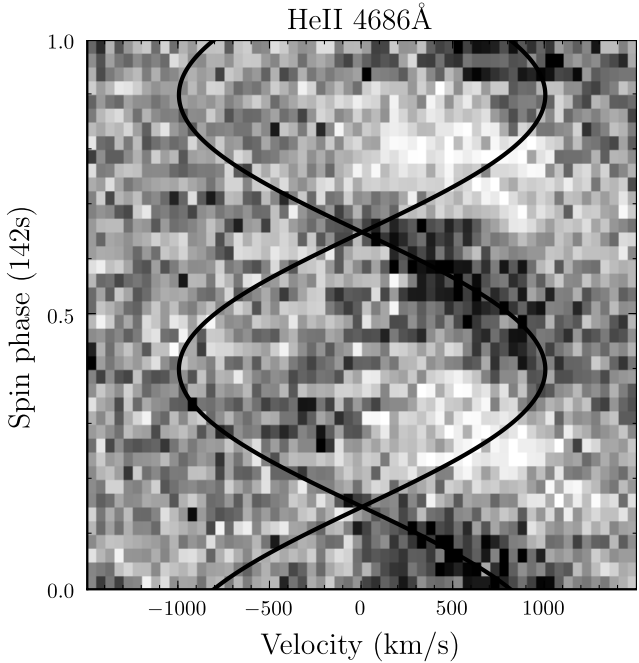}
 \caption{Figure similar to Fig.~\ref{SPIN_4686_71s_rep2} but assuming a spin period of 142\,s. The black curves show the emission that can be expected from inner disc reprocessing of two white dwarf X-ray beams. Two similar regions with enhanced emission can be seen, which suggests that the true spin frequency is rather 71\,s. However, the slope of the sinusoidal curves fits the slope of the enhanced emission better than in the case of 71\,s.}
  \label{SPIN_4686_142s_rep1}
\end{figure}

In Fig.~\ref{SPIN_4686_71s_rep2}, a trail of He~II $\lambda4686$ is plotted after folding on a 71\,s period. The cycle is repeated to ease comparison with the 142\,s trails. Light pixels indicate a line flux that is lower than the average over the spin cycle on that particular wavelength, dark indicates a higher than average flux. The black sinusoidal curve shows the path of the enhanced emission that would be seen when the reprocessing region is a small part of the disc that follows a circular orbit, as also shown on the spin trails of \citet{MartellHorne1995}. Figure~\ref{SPIN_4686_142s_rep1} shows the result of folding on a 142\,s period. For a 71\,s spin period, one reprocessing region is needed. For a 142\,s spin period, one needs two reprocessing regions that are half a spin phase apart to produce the observed periodicity. Therefore, two sine curves with a phase difference of $180^\circ$ have been plotted.

We can confirm, with greater significance, the observation of \citet{MartellHorne1995} that a pulsation is visible in He~II $\lambda4686$ when the beam illuminates the red-shifted part of the back side of the disc, but invisible when the beam points to the front side or the blue-shifted part of the back side. The absence of fluctuations in the light coming from the front side of the disc is not surprising: the inclination is close to 90$^\circ$ and since a disc is not flat but rather concave, our view of the front part might well be blocked by the thick edge (the rim) of the disc. The fact that no fluctuations are seen in the blue-shifted part of the back side of the disc is more puzzling. \citet{MartellHorne1995} propose that the pulsation comes from the threading region, i.e.\ the region at the inner edge of the disc where the material gets trapped by the magnetic field of the white dwarf. If this region is optically thick, a velocity gradient in the direction of our line of sight will result in enhanced emission (\citealt{Horne1995}). 

Another possible explanation is that the X-rays are reprocessed by (optically thick) accretion curtains. Due to the curvature of such curtains, one can imagine that the X-ray beam might only illuminate the curtain side closest to the white dwarf (the `bottom' side), and not the `top' side. If the curtain is optically thick, then it might be the case that we see the bottom side of the curtain when it is receding from us and the non-illuminated top side when it is approaching us. Such a scenario would produce variations in the red-shifted part of the line, but no fluctuations in the blue-shifted part.

A scenario in which the accretion from the disc onto the white dwarf's surface switches on and off during the spin cycle was also considered. This can be the case when the orientation of the WD's magnetic axis is offset from the rotation axis such that the magnetic field disconnects from and reconnects with the material in the disc. One would expect that if at a certain orbital phase the magnetic field is connected when the magnetic pole is red-shifted and disconnected when the pole is blue-shifted, the opposite would be true half an orbit later. Therefore, as much variation should be observed in the blue-shifted parts of the line as in the red-shifted, which is not the case. Furthermore, in Section~\ref{sec_orbphasedep} we will show that the periodicity in the red-shifted wing of the line is observed at all orbital phases, which also rules out this explanation.

Spin trails have been published earlier for IPs with slower spinning white dwarfs, e.g., for Ex Hya in \citet{HellierMason1987}, for FO Aqr in \citet{HellierMason1990}, for AO Psc in \citet{HellierCropper1991}, for BG CMi and PQ Gem in \citet{Hellier1997}, and for V2400 Oph and V1025 Cen in \citet{Hellier1999}. All of these reveal variations in both the blue- and redshifted parts of the lines. One has to keep in mind, however, that the accretion region in DQ~Her might be fundamentally different since its WD is one of the fastest spinning of the known IPs (the spin periods in the other IPs mentioned here range from 805\,s for AO Psc to 4022\,s for Ex Hya). Though technically challenging, high speed spectroscopy of other fast spinning IPs is necessary to check whether DQ~Her's spin effects are common among this group of IPs.


\subsection{71\,s or 142\,s periodicity in He~II $\lambda4686$?}

The spin trail of He~II $\lambda$4686 also potentially offers a means to conclude the discussion on whether the spin period is 71\,s or rather 142\,s with two spots contributing to the pulsation. The slope of the simplistic disc reprocessing model clearly fits the 142\,s trail (Fig.~\ref{SPIN_4686_142s_rep1}) better than the 71\,s trail (Fig.~\ref{SPIN_4686_71s_rep2}). The sinusoidal curves would be fairly correct if a small part of the disc at semi-Keplerian orbits, e.g.,~the inner part of the disc, reprocesses the X-rays of the WD poles. If that is the case, a 142\,s spin period is to be preferred. If the extra emission, however, comes from accretion curtains (as was suggested already for some of the IPs with slower spinning WDs, see e.g.,\ \citealt{HellierMason1990}), a sinusoidal radial velocity trace is too simple and a more detailed model would be required to draw a conclusion. As discussed in Section \ref{sec_DM_HeII}, the Doppler map of this He~II line (Fig.~\ref{DM_4686}) shows a large non-circular component, which can also be explained by the reasoning that the He~II emission would primarily come from an accretion curtain. The fact that we don't see any significant difference between the first half and second half of a 142\,s cycle confirms earlier suggestions that if the spin period is really 142\,s, the two poles would have to be almost identical.


\subsection{Can 71\,s be the beat frequency?}  \label{sec_orbphasedep}

\begin{figure*}
\includegraphics[width=180mm]{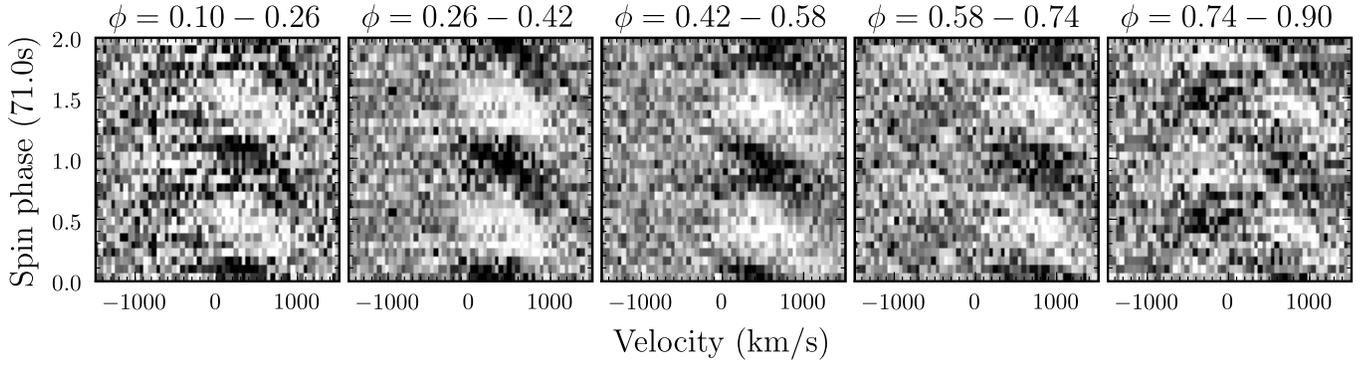}
 \caption{Spin trails of He~II $\lambda4686$ similar to Fig.~\ref{SPIN_4686_71s_rep2}, assuming a 71\,s spin period, but produced using spectra taken during different orbital phase $\left(\phi\right)$ intervals. The variation in the red-shifted wing of the line is visible on all orbital phases, which indicates that the pulsation cannot originate from reprocessing on the bright spot or the secondary star. In that case, the variations would not be visible at late orbital phases, when the bright spot and secondary star are at blue-shifted velocities. A weak variability component is detected in the blue-shifted wing of the line at late orbital phases.}
  \label{SPIN_4686_71s_mphase}
\end{figure*}

\begin{figure*}
\includegraphics[width=180mm]{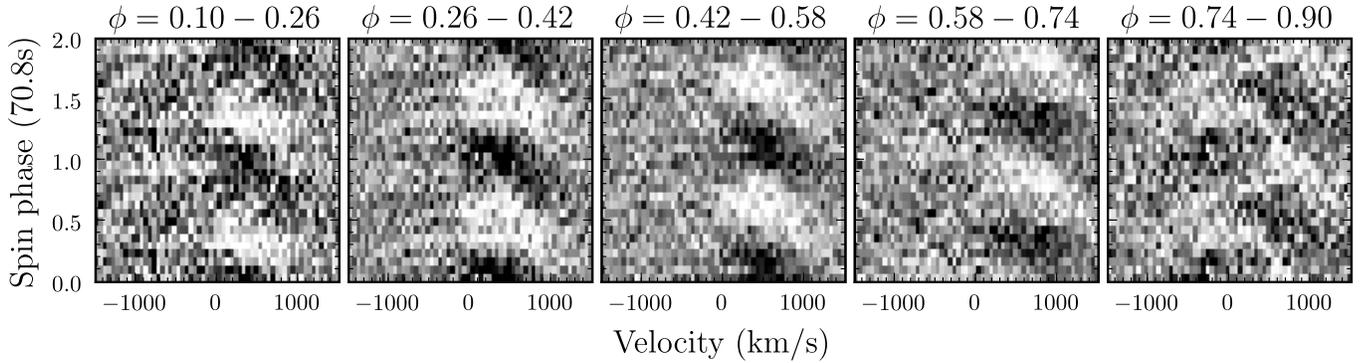}
 \caption{Spin trails of He~II $\lambda4686$ similar for different orbital phase intervals like Fig.~\ref{SPIN_4686_71s_mphase}, but assuming a 70.8\,s spin period. This would be the case if 71\,s is the beat period. The dataset clearly allows to distinguish a 70.8\,s from a 71\,s periodicity, since it can clearly be seen that the pulsation occurs at later spin phases on the plots of later orbital phases. No 70.8\,s periodicity is found, because it is dominated by the 71\,s frequency or because it is absent. }
  \label{SPIN_4686_70s_mphase}
\end{figure*}
Very recently, \citet{SaitoBaptista2009} used the dataset formerly used by \citet{MartellHorne1995} for eclipse mapping. Comparing eclipse maps of different spin phases, they claimed to have found compelling evidence for the existence of a rotating component at the inner radius of the disc, which they associated with an accretion curtain, and pulsating emission from the bright spot in the outer part of the disc. They conclude that mostly the bright spot and to a lesser extent accretion curtains contribute to the 71\,s fluctuations, and that the observed frequency should be the beat frequency instead of the spin frequency because of the larger contribution by the bright spot.

Given the much higher signal-to-noise ratio of our data compared to the spectra used by \citet{MartellHorne1995}, we were able to check the variability in the pulsation of the He~II $\lambda$4686 line over different orbital phases.  Figure \ref{SPIN_4686_71s_mphase} displays the spin trails (folded on the 71\,s period) for five orbital phase intervals. The amplitude of the pulsation in the red-shifted wing of the line seems a little lower at the later orbital phases but the variability is still obvious. This indicates that most of the pulsation originates from the disc, and not from the bright spot or the secondary star. Note that some variability can be seen in the blue-shifted wing of the line at late orbital phases as well.

The 71\,s pulsation in the red-shifted wing at orbital phases where the possible beat frequency contributors (the bright spot and the secondary) are in the blue-shifted part of the line rules out the possibility that the spin period would be 70.8\,s, with the dominant 71\,s pulse being the beat period. The time baseline of the observations is long enough to resolve the difference between a 70.8\,s and a 71.0\,s period. On the 71\,s trails, the pulsation profile in the red-shifted part of the line stays at the same phase, while for 70.8\,s it shifts over 1 spin phase in 1 orbital cycle. This is illustrated on Fig.~\ref{SPIN_4686_70s_mphase} which shows the trails folded on the 70.8\,s period for the same orbital phase chunks as Fig.~\ref{SPIN_4686_71s_mphase}. The shift of the pulsation in the red wing is clearly visible, which also strengthens our believe that if a 70.8\,s periodicity is present at all, it is certainly not dominant over the 71\,s period. 

The detection of the 71\,s pulsation in the red-shifted part of the line at all orbital phases and the fact that eclipse mapping is not as well constrained as Doppler mapping and the spin trails because it only uses data taken during eclipse, during which part of the disc remains hidden from view, we believe that further evidence is necessary to confirm that the He~II-pulsation arises from the curtains. The bright spot might contribute as well, but probably not as much as derived by \citet{SaitoBaptista2009} from the eclipse maps because we can rule out the possibility that 70.8\,s is the true spin period, which would be necessary if the bright spot is the dominant source of the pulsations. 


\subsection{Spin periodicity in the Balmer and He~I lines}

\begin{figure}
\includegraphics[width=84mm]{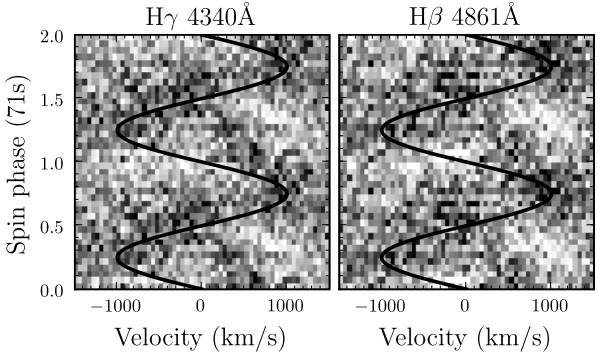}
 \caption{Spin trail similar to Fig.~\ref{SPIN_4340_71s_rep2} for H$\gamma$ and H$\beta$, assuming a 71\,s spin period. Variations are seen in both the blue- and red-shifted parts of the line. The simple disc reprocessing model does not fit the emission features well.}
  \label{SPIN_4340_71s_rep2} \label{SPIN_4861_71s_rep2}
\end{figure}

\begin{figure}
\includegraphics[width=84mm]{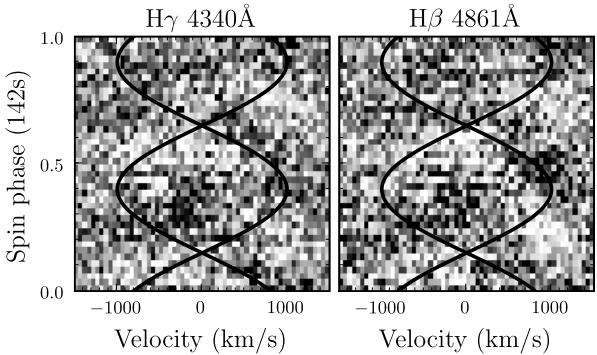}
 \caption{Spin trail similar to Fig.~\ref{SPIN_4861_71s_rep2} but assuming a 142\,s spin period. The disc reprocessing model now fits the slope of the red-shifted emission features well, but fails completely in explaining the variations in the blue-shifted wing of the line.}
  \label{SPIN_4340_142s_rep1} \label{SPIN_4861_142s_rep1}
\end{figure}

Signs of a variation on the spin period are also visible in the trails of the Balmer lines, which are shown in Fig.~\ref{SPIN_4340_71s_rep2} for H$\gamma$ and H$\beta$ folded on the 71\,s period and in Fig.~\ref{SPIN_4861_142s_rep1} for the same lines folded on the 142\,s period. The amplitude is only marginally above noise level, but it seems that there is a fluctuation visible in the blue parts of the Balmer lines as well, which is similar to the variations in He II $\lambda4686$ at late orbital phases. The red wings seem to show a pattern that is different from the one seen in He~II $\lambda$4686. The pulsation is not well fitted by the simple 71\,s pulse model, but the same holds for the 142\,s model. This observation also points in the direction of a more complicated geometry of the reprocessing region. The spin behaviour of H$\alpha$ could not be checked because this line was observed with the red arm of ISIS using exposure times of 15.1\,s, contrary to the other lines which were observed with the blue arm at a 5.1\,s cadence.

The pulsations in the He~I lines are too weak to be detected on spintrails.

\section[]{Conclusions} \label{sec_concl}
We studied high-speed spectroscopy of the prototype intermediate polar DQ~Her. We report the first detection of spiral arms in the accretion disc of an intermediate polar. Spiral arms are seen in Doppler maps of H$\gamma$, He~I $\lambda4472$ and He~I $\lambda6678$. Spirals arise from tidal effects from the secondary star on the accretion disc. They were found before in dwarf novae in outburst and in nova-like variables. Since spiral shocks are believed to be an additional mechanism for the transport of angular momentum in a disc besides viscosity effects, they can form a crucial clue in the understanding of disc related accretion processes and other astrophysical phenomena like planet and star formation. If spiral arms can propagate far enough into the disc, as indicated by the simulations of \citet{MurrayArmitage1999}, they can modulate the accretion rate onto the white dwarf. Inner disc asymetries such as those produced by spiral arms can therefore possibly explain the sidebands of the spin frequency that are often observed in X-ray and optical light curves of IPs. Until now, it was believed that these sidebands indicate that (part of) the accretion proceeds via a direct stream from the first Lagrangian point to the white dwarf rather than through the disc. If spiral arms in accretion discs can explain the sidebands equally well, accretion streams might be less frequent than currently thought.

We built upon the study of the pulsation pattern in emission lines from the accretion disc as presented earlier by \citet{MartellHorne1995}. We can confirm with greater significance that reprocessed light from the WD beam is mainly visible in He~II $\lambda$4686 when the beam points to the red-shifted back side of the disc. The slope of the pulsation on a spin trail does not match with what would be expected from a simple model for a spin period of 71\,s in which a small region of the disc reprocesses the X-rays from the WD beam. The same model for two reprocessing spots and a 142\,s spin period results in a better fit. However, the Doppler map of He~II $\lambda$4686 suggests that the line emitting region has a substantial non-Keplerian component. Reprocessing in accretion curtains might be a solution to fit both the slope of the pattern and the absence of observed pulsations in the blue-shifted part of the line, but further modelling is required to check this suggestion. The lack of a good model thus currently prevents us from settling the discussion as to whether 71\,s or 142\,s is the true white dwarf spin period. The spin trails of the He~I and Balmer lines show weak pulsation components as well, with puzzling slopes. We are not aware of any fully developed IP model that can quantitatively explain our observations. The spin trails clearly provide much more information than can be derived from photometric studies and provide a strong test for enhanced IP accretion models in the future. 

Pulsations are still observed at a 71\,s periodicity in the red-shifted wing of He~II $\lambda$4686 at orbital phases where the bright spot is in the blue-shifted part of the disc. We therefore refute the assertion by \citet{SaitoBaptista2009} that most of the reprocessed light comes from the bright spot, and rule out the possibility that 70.8\,s would be the true spin period.

\subsection*{Acknowledgments}
The authors wish to thank Keith Horne and Koji Mukai for interesting discussions on the topic. 

SB wishes to thank Conny Aerts, Jos Rogiers, Boris G\"ansicke and his co-authors for their efforts to arrange his Erasmus stay at the University of Warwick, which led to the research presented in this article.

This research is based on observations made with the William Herschel Telescope operated on the island of La Palma by the Isaac Newton Group in the Spanish Observatorio del Roque de los Muchachos of the Instituto de Astrof'sica de Canarias.

The research leading to these results has received funding from the
European
Research Council under the European Community's Seventh Framework Programme
(FP7/2007--2013)/ERC grant agreement n$^\circ$227224 (PROSPERITY), as
well as
from the Research Council of K.U.Leuven grant agreement GOA/2008/04.
During this research TRM and DS were supported under grants from the UK's 
Science and Technology Facilities Council (STFC, ST/F002599/1 and 
PP/D005914/1).

\label{lastpage}

\end{document}